\documentstyle[aps]{revtex} 
 \begin{document}
 \author{C. S. Unnikrishnan$^1$ and Shomeek Mukhopadhyay$^2$}
 \address{$^1${\em Gravitation experiments Group}, $^2$ {\em Theoretical Physics Group}%
 , \\{\em Tata Institute of Fundamental Research, Homi Bhabha Road, Mumbai -}%
 \\400 005, India}
 \title{Comments on `Sonoluminescence as Quantum Vacuum Radiation'}
 \date{May 1996}
 \maketitle
 
 \begin{abstract}
 We argue that the available experimental data is not compatible with models
 of sonoluminescence which invoke dynamics of the interface without regard to
 the compositional properties of the trapped gas inside the bubble. The main
 point is that while the actual sonoluminescence intensity is sensitive to
 the minute details of the composition of the trapped gas, like trace amount
 of inert gases like Argon, and the temperature of the surrounding water, the
 vacuum radiation models rely entirely on the dynamics of the interface and
 therefore do not predict the observed dependencies in experiments. The
 spectral features also are dependent on these parameters whereas the vacuum
 radiation models do not incorporate such dependencies. \medskip\ 
 
 
 \medskip\ 
 \end{abstract}
 
 Recently there have been some attempts to model the phenomenon of single
 bubble sonoluminescence as arising from the modification of the
 electromagnetic vacuum \cite{swing,claud}. Earlier Schwinger has suggested
 that the origin of sonoluminescence might be in the dynamic Casimir effect
 generated by the moving bubble - liquid interface \cite{swing}. This has
 inspired many other studies and recently Claudia Eberlien has suggested \cite
 {claud} that the radiation in sonoluminescence may be generated due to an
 effect similar to the Unruh effect, where the acclerations of the dielectric
 interface is responsible for the thermal like radiation. Though the
 parameter values needed to fit the general characteristics of thermal like
 spectrum at the observed intensity are probably slightly unrealistic in
 these models, they are interesting and, in the absence of any realistic good
 models of sonoluminescence so far, they should be taken with some
 seriousness. But, as we will argue now, there is already enough experimental
 facts available which make these models unrealistic. In fact according to
 us, any model which invoke only the dynamics and properties of the
 bubble-liquid interface and not the details of the trapped gas and the
 environment like ambient temperature are missing out some very important
 observations on sonoluminescence and therefore are not viable.
 
 These models seems to ignore the important experimental fact that no
 sonoluminescence is observed in situations where there is no traces of inert
 gases in the trapped air \cite{put1} or that the intensity drops drastically
 when the ambient temperature is increased by a mere 5\% \cite{put2}. Also,
 the spectrum is affected significantly by such temperature changes. In fact
 attempts to generate sonoluminescence with various trapped gases and other
 liquids have conclusively shown that the composition of the air in the
 bubble is an extremely important factor \cite{put1,ara}. Changes in this
 composition can drastically affect the intensity of radiation and even stop
 the emission. {\em Lack of noble gases completely stops the emission and
 addition of even 1\% of Argon, Xenon or Helium is sufficient for
 sonoluminescence to occur, though there is no change in the refractive index
 significant enough to affect the expressions derived in the vacuum radiation
 models}. This probably is the most crucial test of such models since the
 difference in refractive indices at the interface is what defines the
 `moving mirror', and we think that these observations have conclusively
 ruled out the possibility that the vacuum radiation is the dominant
 component of sonoluminescence (We are not contesting the possibility that
 there may be a small component in the emission coming from such exotic
 phenomena).
 
 The field theory calculation has no dependence on the small ambient
 temperature in the experimental situation, and certainly is not sensitive to
 the small changes in this temperature. This is expected in such models since
 the only important feature in these models to generate radiation is the
 dynamics of the interface and small temperature changes negligible compared
 to effective temperatures generated by the moving interface is insignificant
 in these models. But the experimental results have a sensitive dependence on
 the ambient temperature and this is a further test of the vacuum radiation
 models and again the models do not adequately explain the observed features.
 
 The point is that, the experiments have shown that there are situations
 without any detectable radiation, almost as a rule, in which there is stable
 bubble with its interface undergoing the same dynamics as in the case when
 there is radiation. The difference is not in the dynamics of the interface,
 but in the minute details of the composition of the trapped gas or small
 changes in the ambient temperature.
 
 There may be other such features which are incompatible with vacuum
 radiation models and it is worth analysing these models in the light of {\em %
 all these experimentally verified features }(There have been recent attempts
 to do a more detailed modelling of sonoluminescence as quantum vacuum
 radiation, taking into account of some of the experimentally observed
 features \cite{claud2}). With such an approach, it may be even possible
 refine the vacuum radiation models, for example to incorporate some
 dependence of the velocity turnaround time on the amount of dissolved gases
 and thereby to obtain some dependence of the emission on such parameters
 even in these models.
 
 \medskip\ 
 
 \noindent Electronic address:
 
 \noindent $^1$unni@tifrvax.tifr.res.in, $^2$shomeek@theory.tifr.res.in

 \end{document}